\documentclass[aps,prl,twocolumn,showpacs,floatfix]{revtex4-1}
\usepackage{graphicx}
\usepackage{color}
\usepackage{amsmath}
\usepackage{amssymb}
\usepackage{amsfonts}
\usepackage{bm}
\definecolor{scarred}{rgb}{0.75,0.0,0.0}
\usepackage{hyperref}
\hypersetup{
colorlinks=true,final=true,
        linkcolor=blue,
        citecolor=blue,
        filecolor=blue,
        urlcolor=blue,
}

\begin{document}


\title{Ultra-fast electric field controlled spin-fluctuations in the Hubbard model}

\author{Nagamalleswararao Dasari}\email{nagamalleswararao.d@gmail.com}
\author{Martin Eckstein}\email{martin.eckstein@fau.de}
\affiliation{Department of Physics, University of Erlangen-Nuremberg, 91058 Erlangen, Germany}
\begin{abstract}
Highly intense electric field pulses can move the electronic momentum occupation in correlated metals over large portions of the Brillouin zone, leading to phenomena such as dynamic Bloch oscillations. Using the non-equilibrium fluctuation-exchange approximation for the two-dimensional Hubbard model, we study how such non-thermal electron-distributions drive collective spin and charge fluctuations. Suitable pulses can induce a highly anisotropic modification of the occupied momenta, and the corresponding spin dynamics results in a transient change from antiferromagnetic to anisotropic ferromagnetic correlations. To good approximation this behavior is understood in terms of an instantaneous response of the spin fluctuations to the single-particle properties, opposite to the conventional time-scale separation between spin and electron dynamics.
\end{abstract} 
 
\maketitle 

Ultra-short and highly intense laser pulses have opened novel pathways to control quantum materials \cite{Basov2017,Giannetti2016}.  In this context, a large body of theoretical and experimental work aims to understand the electronic {\em single-particle} properties out of equilibrium. There are  many detailed accounts of the ultrafast momentum and  energy-resolved electron dynamics, including recent studies of strongly correlated quasi-particles or band electrons in two-dimensional materials \cite{Gierz2015,Bertoni2016,Rameau2016}. Moreover, non-perturbative electric fields can coherently drive electrons  over large portions of the Brillouin zone, thus enabling Floquet band-engineering \cite{Wang2013,McIver2018} or the observation of Bloch oscillations \cite{Schubert2014} and Zener tunnelling \cite{Higuchi2017} in solids. On the other hand, many rich properties of correlated materials arise from the interplay of the electronic structure with charge, spin, and orbital fluctuations, and  intriguing pathways for transient light-induced or enhanced orders have been theoretically proposed \cite{Sentef2016, Knap2016, Murakami2017,Nava2017,Li2018}. This poses the immediate question whether already the above mentioned single-particle dynamics can imply a nontrivial collective response, such as, e.g., signatures of  spin dynamics during Bloch oscillations in moderately  correlated systems. In insulators (or Mott insulators), thermalization of electrons is slowed down by the gap so that  non-thermal electron populations may live long enough to induce ``hidden states'' which differ from any equilibrium phase \cite{Li2018}, or laser driving can directly induce non-thermal spin-wave populations \cite{Walldorf2018}. In metallic systems, in contrast, electron thermalization is typically considered to be one of the fastest timescales, both at weak interactions, where standard kinetic equations apply \cite{HaugBook}, and for strongly correlated electron liquids close to the Mott transition without well-defined quasiparticles \cite{Eckstein2011,Ligges2018}. Assuming that electrons are thermalized on timescales relevant for the collective dynamics is therefore often an excellent approximation, which forms the basis for phenomenological two-temperature models \cite{Allen1987} and intriguing predictions from the non-equilibrium field theory \cite{Lemonik2017,Lemonik2018}. However, while a clear separation of timescales holds certainly true for the long-wavelength fluctuations close to critical points, fluctuations at shorter scales may still be fast enough to display a nontrivial dependence on nonthermal electron populations.  In the present work, we investigate this issue by studying the dynamics of electrons and spin in the two-dimensional Hubbard model. 

The Hubbard Hamiltonian is given by
\begin{align}
H 
=
-\!\!\!\!\sum_{\langle \bm R,\bm R' \rangle,\sigma}
\!\!t_{\bm R-\bm R'}\,
c_{\bm R\sigma}^\dagger
c_{\bm R'\sigma}^{\phantom \dagger}
+
U
\sum_{\bm R}
n_{\bm R,\uparrow}
n_{\bm R,\downarrow},
\label{THEmodel}
\end{align}
where $c_{\bm R\sigma}^\dagger$ creates an electron with spin $\sigma\in\{\uparrow,\downarrow\}$ on 
site $\bm R$ of a square lattice of size $L^2$, $U$ is the repulsive on-site interaction; $t_{\bm R-\bm R'}$ is the nearest neighbour hopping, corresponding to a dispersion $\epsilon_{\bm k}(t) = \epsilon_0(\bm k-\bm A(t))$ with $\epsilon_0(\bm k) = -2t_\text{hop}[\cos(k_x a) + \cos(k_y a)]$. Here the electric field $\bm E(t)$ is incorporated using the Peierls substitution, with the vector potential $\bm A(t)$ and $\bm E(t)=-\partial_t \bm A(t)$. We choose units  $a=1$, $e=1$, and $\hbar=1$. The tunnelling matrix element $t_{\text{hop}}=1$ sets the energy scale. 

The non-equilibrium dynamics of the model is discussed within the Keldysh formalism on the $L$-shaped time contour $\mathcal{C}$, suited to study the dynamics of a system which is initially in thermal equilibrium at a given temperature $T$ \cite{footnote01}.
We study the dynamics in terms of the contour-ordered electronic Green's functions
$G_{\bm R-\bm R'}(t,t') 
=
-i 
\langle
T_{\mathcal{C}} c_{\bm R}^{\phantom \dagger}(t) c_{\bm R'}^\dagger (t')
\rangle
$,
and the collective propagator
$\chi^{\alpha}_{\bm R-\bm R'}(t,t') 
=
-i 
\langle
T_{\mathcal{C}} \hat X^{\alpha}_{\bm R}(t) \hat X^{\alpha}_{\bm R'} (t')
\rangle$,
where $\hat X^{\alpha}$ can be spin  $S_{\bm R}=\sum_{\sigma\sigma'} c_{\bm R\sigma}^\dagger \tau_{\sigma\sigma'} c_{\bm R\sigma'}^{\phantom \dagger}$ ($\alpha\equiv s$), or charge $X^{\alpha}_{\bm R}=\sum_{\sigma} n_{\bm R\sigma}$ ($\alpha\equiv c$).  The spatial Fourier transform is defined as $f_{\bm R} = \frac{1}{L^2}\sum_{\bm q} e^{i\bm q\bm R} f_{\bm q}$.  From the contour-ordered functions we obtain time-dependent spectra (see below), the gauge-invariant momentum distribution $\tilde n_{\bm k}(t) = \langle c_{\bm k-\bm A(t)}^\dagger (t) c_{\bm k-\bm A(t)}^{\phantom \dagger}(t)\rangle =-iG^<_{\bm k-\bm A(t)}(t,t)$, and the spin and charge-correlations $C^{\alpha}_{\bm q}(t)= \langle\hat X^{\alpha}_{\bm q}(t) \hat X^{\alpha}_{-\bm q} (t) \rangle=i\chi^{\alpha<}_{\bm q}(t,t)$.

To study the interplay of electrons and collective fluctuations at moderately strong $U$, we employ the fluctuation-exchange (FLEX) approximation \cite{Bickers1989}, a $\Phi$-derivable (i.e., energy and number-conserving) approximation designed to treat the interaction of electrons with charge, spin, pairing, or  orbital fluctuation channels. The formulation of the diagrammatic approach \cite{Bickers1989b} is identical on the Matsubara and on the Keldysh time-contour. The  approximation for the collective propagators is given by the RPA series
\begin{align}
\chi^{\alpha}_{\bm q} = (1-U_{\alpha}\Pi_{\bm q})^{-1} \ast \Pi_{\bm q}.
\label{rpa}
\end{align}
Here $\ast $ denotes a convolution in contour-time \cite{footnote01}, the interaction is $U_c=-U_s=U$ for charge and spin, respectively, and $\Pi_{\bm R}(t,t') =-i G_{\bm R}(t,t') G_{-\bm R}(t',t)$ is the bare susceptibility, which is identical for charge and spin in the paramagnetic phase. The electron self-energy is given by the second-order diagram, supplemented by the contributions from the fluctuation self-energy $\Sigma^{\alpha}_{\bm R}(t,t') = i G^{\alpha}_{\bm R}(t,t') F^{\alpha}_{\bm R}(t,t')$, $F_{\bm R}(t,t')=U^2 \chi^{\alpha}_{\bm R} (t,t')$ beyond second order. The Keldysh FLEX has been used to study momentum-dependent quasiparticle relaxation \cite{Sayyad2018} and transient Floquet engineering \cite{Dasari2018}. Here we focus on the metallic phase in the repulsive Hubbard model at half filling, where magnetic fluctuations are dominant, and therefore include only the magnetic fluctuation channel into the self-energy. The numerical simulations are performed on a finite grid of $L^2$ momenta  ($L=28$ for the results below, which is sufficient to obtain converged results for the short-range correlations under investigation). The numerical implementation is based on the libCNTR non-equilibrium Green's functions library \cite{footnote02}.

\begin{figure}[tbp]
\includegraphics[width=\columnwidth]{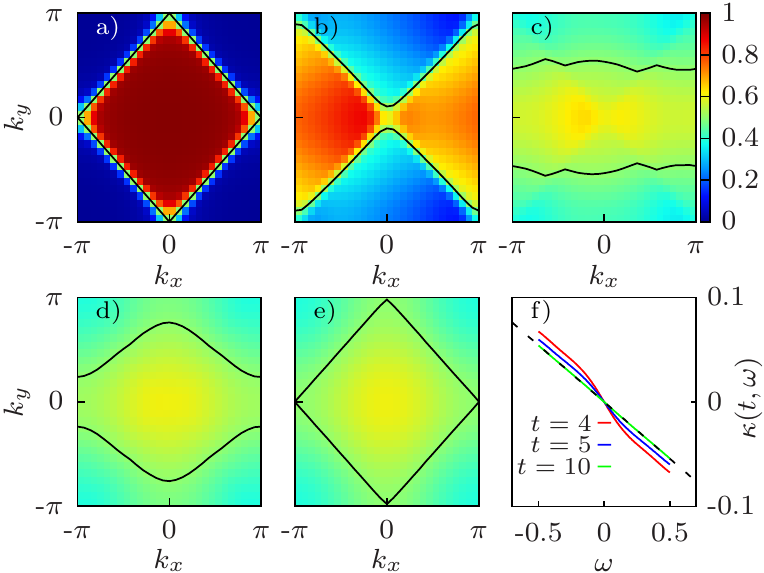}
\caption{Evolution of the momentum distribution $\tilde n_{\bm k}(t)$ for $U=1.5$ and a half-cycle pulse with momentum transfer  $\int dt \bm E=(\pi,0)$, for times $t=0$ before the pulse (a), $t=4$ directly after the pulse (b), and  $t=9.4,14.6,30.0$ during the relaxation towards the hot-electron state (c)-(e). The solid lines show the surface defined by the occupation $\tilde n_{\bm k}=0.5$. (f) Logarithmic ratio $\kappa(t,\omega)=\ln [G^<(\omega,t)/G^>(\omega,t)]$, which tends to $\kappa(t,\omega)=-\beta_* \omega$ ($\beta_*=0.11$, dashed line) in a thermal state with temperature $T_*=1/\beta_*$.}
\label{fig1}
\end{figure}

{\em Results -- Single-particle properties:}
In Fig.~\ref{fig1}a)-e) we exemplarily show the momentum occupation $\tilde n_{\bm k}(t)$ for a moderately correlated system ($U=1.5$), which is driven by a half-cycle pulse of the form $\bm E(t)= \bm \eta (A_{0}\pi/2t_0) \sin(\pi t/t_0)$ (for $0<t<t_0$) with polarization $\bm \eta=(1,0)$. Without interaction $U$, the pulse would simply shift the electrons by the momentum transfer $\Delta\bm  k = \int dt \bm E\equiv \bm A_0 $ [$\bm A_0=(\pi,0)$ in Fig.~\ref{fig1}]. In the interacting case we observe a similar shift if the pulse is not too long, up to a broadening of the distribution by electron-electron scattering. Subsequently, electrons relax back to the band minimum and finally thermalize at elevated temperature. The kinetic energy $e_{kin}$ is roughly zero after the pulse, because the distribution is symmetrically centred around $\bm k=(\pi,0)$. While $e_{kin}=0$ would correspond to infinite temperature, during the relaxation the interaction energy is increased and $e_{kin}$ is decreased, such that the final temperature $T_*$ is of the order of the bandwidth. The thermalization can be confirmed also from the dynamic response functions: Figure \ref{fig1}f) shows the logarithmic ratio $\kappa(t,\omega)=\ln [G^<(\omega,t)/G^>(\omega,t)]$ for the local Green's function $G$. (Time-dependent spectra are obtained from the partial Fourier transform $G^{>,<}(t,\omega) = \pm\text{Im} \int ds \,G^{>,<}(t,t-s) e^{i\omega s}$.) The linear relation $\kappa(\omega)=-\omega/T_*$, which is reached for long times, proves that the fluctuation-dissipation theorem is satisfied, such that local single-particle quantities can be considered in thermal equilibrium. 

To characterize the dynamics it is interesting to look at the ``Fermi-surface'' defined by $\tilde n_{\bm k}=0.5$, although for a general non-equilibrium state, this surface neither corresponds to a maximum of the quasiparticle scattering time, nor to a discontinuity in $\tilde n_{\bm k}$. (Note that after interaction quenches from an initial state at $T=0$ there remains an {\em exact} discontinuity in the momentum occupation for some time \cite{Moeckel2008,Eckstein2009,Uhrig2009}, whose shape is however not renormalized.) During the time-evolution the surface $\tilde n_{\bm k}=0.5$ changes  from a closed to an open, quasi one-dimensional topology. Because collective excitations and their instabilities strongly depend on occupied states, this already suggests that the relaxation can have a strong effect on the two-particle fluctuations.

\begin{figure}[tbp]
\includegraphics[width=\columnwidth]{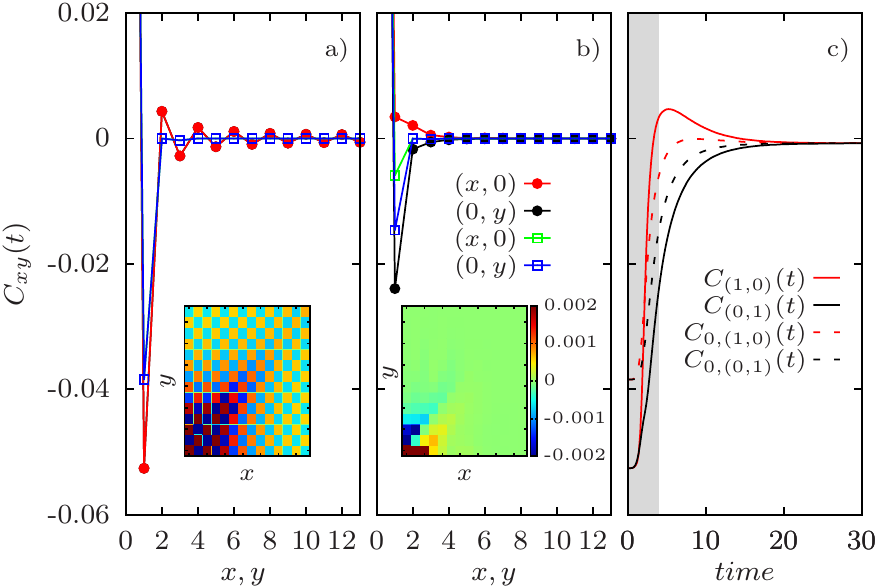}
\caption{Spin correlation function $C^{(s)}_{\bm R}(t)$ along the $x$ and $y$-directions for the same parameters as Fig.~\ref{fig1}, at time $t=0$ (a) and $t=4$ (b). At $t=0$, the system is still isotropic, $x\equiv y$. Open symbols show the correlations obtained from the bare susceptibility, $C_{0,\bm R}(t)$. The inset shows  $C^{(s)}_{\bm R}(t)$ in false color. (c) Time-evolution of the nearest neighbour correlations $C_{(1,0)}$ and $C_{(0,1)}$ in $x$ and $y$ directions, obtained from the full (solid lines) and bare (dashed lines) spin correlation function. The shaded area indicates the duration of the pulse.}
\label{fig2}
\end{figure}

Before discussing this two-particle physics, we briefly comment on the half-cycle pulse. While $\int dt \,\bm E=0$ for  a conventional electromagnetic pulse, the half-cycle pulse with $\int dt \bm E\neq0$ allows us to study in a simple manner both the coherent dynamics during the application of a strong field and the relaxation dynamics in the absence of a field, which can both be accessed in different experimental settings. Furthermore, an asymmetric pulse with $\int dt \,\bm E=0$, consisting of an intense first half cycle followed by a longer and weaker second half, would lead to a similar evolution of the single-particle occupations as in Fig.~\ref{fig1}. Such asymmetric pulses have been proposed to engineer distributions \cite{Tsuji2012}, and can  lead to a population inversion if the pulse is polarized along the $(11)$-direction.

{\em Results -- Spin-correlations:} 
Figure~\ref{fig2} shows the real-space spin correlation function $C^{(s)}_{\bm R}(t) = \langle S^z_{\bm R}(t) S^z_{\bm 0}(t)\rangle$ for the same set of parameters as Fig.~\ref{fig1}. The correlations evolve from short-range anti-ferromagnetism in the initial state to a strongly thermally suppressed anti-ferromagnetism in the final state. During the evolution, however, we observe an entirely different pattern, with ferromagnetic correlations along the $x$-direction (see Fig.~\ref{fig2}c for the nearest neighbour correlations along the $x$ and $y$ axis). Importantly, one can see that the correlations obtained from the bare susceptibility, $C_{0,\bm R}(t)=i\Pi_{\bm R}^<(t,t)$, which reflect the statistical correlations of independent electrons, remain antiferromagnetic in all directions throughout the evolution. The reversal of the spin-correlations thus happens as a consequence of the collective response.

\begin{figure}[tbp]
\includegraphics[width=\columnwidth]{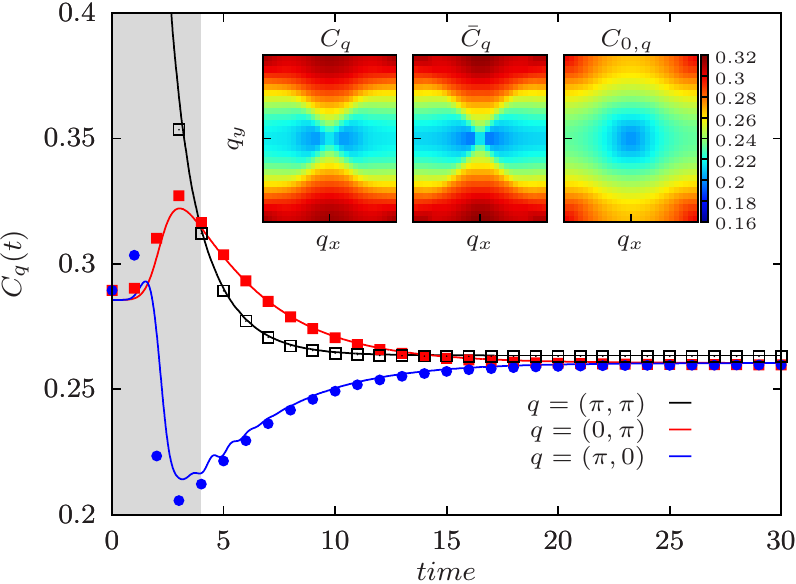}
\caption{(a) Correlation $C^{(s)}_{\bm q}(t)$ (lines) and effective steady-state $\bar C^{(s)}_{\bm q}(t)$ (dots with the same color) for three momenta, and the same parameters as Fig.~\ref{fig1}. The shaded area indicates the duration of the pulse. The inset shows $C^{(s)}_{\bm q}$, $\bar C^{(s)}_{\bm q}$, and $C^{(s)}_{0,\bm q}$ in the full Brillouin zone at time $t=4$. }
\label{fig3}
\end{figure}

In general, the change of the electronic occupation modifies the effective action for the collective modes, and thus inflicts a time-dependent force to drive their dynamics. The numerical results show that at least some part of the spin fluctuations respond faster than the electron thermalization. It is thus interesting to test a scenario which is precisely opposite to the conventional adiabatic separation between fast electrons and slow spin, and in which the spin correlations instead follow the electron dynamics in a quasi-instantaneous manner. As we show below, the numerical results indeed support the latter scenario, as one can rather accurately reconstruct the spin correlations $C^{(s)}_{\bm q}(t)$  from a non-equilibrium steady state that is determined by the electron distribution $\tilde n_{\bm k}(t)$ at the same time $t$:

We start from $\Pi_{\bm q}(t,t')$, which determines the correlation function of the collective modes through the RPA equation \eqref{rpa}. Because the system is weakly interacting, a first-order approximation $\bar \Pi_{\bm  q}(\omega;t)$ for $\Pi_{\bm q}(t,\omega)$ is obtained from the bare response of independent electrons in a non-equilibrium steady state with momentum occupations $\bar n_{\bm k}= \tilde n_{\bm k}(t)$. Here and in the following, barred quantities like $\bar n$ and $\bar \Pi$ correspond to the non-equilibrium steady state, which depends on time only parametrically. $\bar \Pi_{\bm  q}(\omega;t)$  is just given by the  Lindhardt expression $\bar \Pi^<_{\bm q}(\omega;t) = \frac{1}{L^2}\sum_{\bm k} \bar n_{\bm k}(t)(1-\bar n_{\bm k-\bm q}(t))\delta(\epsilon_{\bm k}-\epsilon_{\bm k-\bm q}-\omega)=\bar \Pi^>_{-\bm q}(-\omega;t) $. One can then evaluate  Eq.~\eqref{rpa} in the $2\times2$ Keldysh matrix representation, using $\bar \Pi_{\bm q}(\omega)$ as an input to obtain a non-equilibrium steady state result  $\bar \chi_{\bm q}(\omega;t)$, and thus the steady state correlations $\bar C_{\bm q}(t)= \frac{1}{2\pi i}\int d\omega \,\bar \chi^{<}_{\bm q}(\omega;t)$. Figure \ref{fig3} (inset) shows that the qualitative structure of $C^{(s)}_{\bm q}(t)$  in the transient state is reproduced by the effective steady state $\bar C^{(s)}_{\bm q}(t)$. Again we emphasize that both $\bar C^{(s)}_{\bm q}(t)$ and $C^{(s)}_{\bm q}(t)$ differ from the bare response $C_{0,\bm q}(t)=i\Pi_{\bm q}^<(t,t)$, which retains its maximum at the antiferromagnetic point $\bm q=(\pi,\pi)$, while the collective response develops a maximum at $(0,\pi)$. The lines in the main panel of Fig.~\ref{fig3} show that the comparison is quantitatively accurate for the characteristic momenta $\bm q=(\pi,\pi),(\pi,0),(0,\pi)$ for all times, which confirms the quasi-instantaneous response of the spin to the electrons. In accordance with the real-space picture (Fig.~\ref{fig2}c), the antiferromagnetic correlations at  $\bm q=(\pi,\pi)$ get strongly suppressed ($C^{(s)}_{\bm q}\approx0.25$ corresponds to a featureless high-temperature state), while correlations along the $q_x$ and $q_y$ axis are enhanced and reversed, respectively.

\begin{figure}[tbp]
\includegraphics[width=\columnwidth]{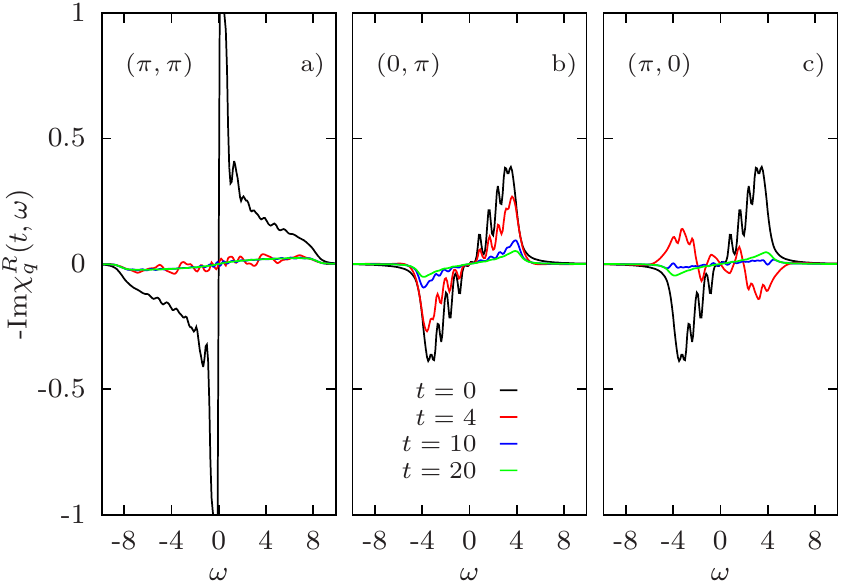}
\caption{Spectra of spin fluctuations for the same excitation as in Fig.~\ref{fig1}, at $\bm q=(\pi,\pi)$ (a), $\bm q=(0,\pi)$ (b), and $\bm q=(\pi,0)$ (c), for initial time $t=0$, directly after the pulse $t=4$, and during the relaxation. Note that at $\bm q=(\pi,0)$ negative spectral weight is formed as a consequence of the electronic population inversion in this direction.}
\label{fig4}
\end{figure}

The fast response of the spin can be further explained by looking at the spectral function $-\frac{1}{\pi} \text{Im} \chi^R_{\bm q}(t,\omega)$ of the collective modes (Fig.~\ref{fig4}). The latter is obtained by partial Fourier transform $\chi_{\bm q}^{R}(t,\omega) = \int_0^{s_\text{max}} ds \,e^{i\omega s}\,\chi_{\bm q}^{R}(t,t-s) $, where the superscript $R$ denotes the retarded component, and $s_\text{max}$ is set by the simulated time. An {\em exact} interpretation of the RPA equation \eqref{rpa} is that a collective field with response function $\chi^R$ is driven by a stochastic force due to electronic quantum and thermal fluctuations with autocorrelations proportional to $i\Pi_{\bm q}^<(t,t')$ \cite{footnote03,KamenevLecture}. Hence, if both the response time set by $\chi^R$ and the autocorrelation time set by the noise is of the order of the bandwidth, the spin fluctuations can follow the single-electron state on the inverse hopping time. In the initial state at $t=0$, slow modes at $\bm q=(\pi,\pi)$ exist because of the vicinity of the antiferromagnetic instability (narrow peak close to $\omega=0$ in Fig.~\ref{fig4}a). These features are however quickly suppressed with time, leading to response with a spectral width of the order of the bandwidth, i.e., few inverse hoppings in the time domain. 

\begin{figure}[tbp]
\includegraphics[width=\columnwidth]{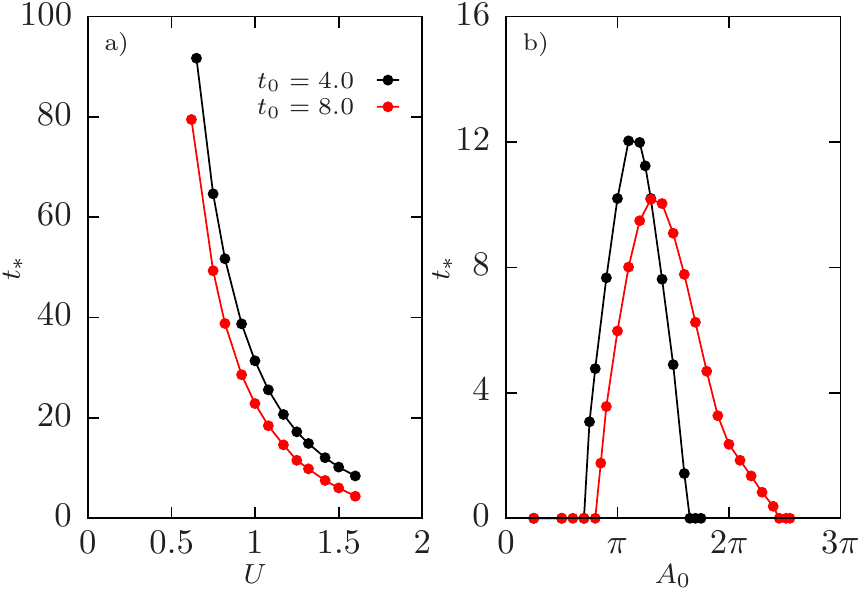}
\caption{
The duration of reversal of spin correlations for different values of $U$ and pulse durations $t_0$ (a), and for different pulse amplitudes $A_0$ at  $U=1.5$ (b).}
\label{fig5}
\end{figure}

The spin-response on the tunnelling timescale indicates that the anti-adiabatic behavior of the short-range spin fluctuations will  become more accurate towards weaker interactions, because electron thermalization slows down like $U^{-2}$. (The absolute value of the collective response of course decreases in this limit). We have performed simulations for a wide range of interactions, different pulse amplitudes $\bm A_0\equiv(A_0,0)$ and pulse durations $t_0$, and found that the reversal of spin correlations is indeed a rather robust feature: Figure~\ref{fig5} shows the duration $t_*$ of the time interval where reversal $C^{(s)}_{(\pi,\pi)}(t)<C^{(s)}_{(0,\pi)}(t)$ is observed, which increases with decreasing $U$ (Fig.~\ref{fig5}a) in agreement with the argument above. Even a full cycle $A_0=2\pi$ can reverse the correlations (Fig.~\ref{fig5}b). Finally, simulations confirm that also the charge response of the system is rapid, but  without significant features in the repulsive Hubbard model at half-filling.

In conclusion, we have shown that non-thermal electrons can quasi-instantaneously drive a non-trivial spin response in a  correlated metal. This opens the intriguing possibility to observe collective electron dynamics driven by ultra-strong THz fields on the sub-cycle timescale, similar to the observation of sub-cycle dynamics on the single-particle level \cite{Schubert2014}. It will be interesting to possibly enhance this collective dynamics by driving electrons dynamically through van-Hove singularities or flat-band regions.  Though challenging, the collective physics might be accessible with time-resolved electron-energy-loss spectroscopy (for the charge dynamics), or  using noise-correlations in time-resolved photoemission, which should be accessible through state-of-the-art momentum microscopes \cite{Stahl2018}. Furthermore, the collective spin response can be also be measured  in cold atoms, where both short-range spin correlations \cite{Greif2015} and coherent motion of particles in the Brillouin zone have been observed.

We acknowledge discussions with J.~Li, A.~Lichtenstein, E.~Stepanov, Ch.~Stahl, and Ph.~Werner. This work was supported by the ERC starting grant No.~716648. The calculations have been done at the RRZE of the University Erlangen-Nuremberg.


\begin{thebibliography}{34}
\bibitem{footnote03}
In order to explain this notion of a noise-driven spin-dynamics, it is convenient to resort to the Hubbard-Stratonovich field $\phi_{\bm q}(t)$ with propagator $F_{\bm q}(t,t')=[U_s^{-1}-\Pi_{\bm q}]^{-1}=U_s\chi_{\bm q}^{(s)}U_s+U_s$, whose fluctuations $F_{\bm q}^<=U_s^2\chi_{\bm q}^<$ are directly  related to the spin. Following Langreth rules, the equation for the lesser component of the field is $F_{\bm q}^<= F_{\bm q}^R \ast \Pi_{\bm q}^< \ast F^A_{\bm q}$. This equation is equivalently represented by $iF_{\bm q}^<(t,t')=\langle  \phi_{-\bm q}(t') \phi_{\bm q}(t) \rangle_\xi$, where $\phi_{\bm q}$ responds to a time-dependent force $ \xi_{\bm q}(t)$ according to the equation  $\phi_{\bm q}(t)=\int_{-\infty }^t dt' F^{R}_{\bm q}(t,t') \xi_{\bm q}(t')$, and the average $\langle\cdots \rangle_\xi$ is taken over a stochastic force with autocorrelation $ \langle \xi_{-\bm q}(t')  \xi_{\bm q}(t)\rangle_{\xi} =i\Pi_{\bm q}^<(t,t')$.
\end{thebibliography}

%

\end{document}